\documentclass[a4paper,11pt]{article}
\pdfoutput=1 

\usepackage{jinstpub} 
\usepackage{lineno}

\title{\boldmath Simulation results for a low energy nuclear recoil yields measurement in liquid xenon using the MiX detector}

\author[a,1]{C.S. Amarasinghe,\note{Corresponding author.}}
\author[a,2]{R. Coronel,\note{Currently at Stanford University.}}
\author[a]{D.Q. Huang,}
\author[a]{Y. Liu,}
\author[a,3]{M. Arthurs,\note{Currently at SLAC National Accelerator Laboratory.}}
\author[a]{S. Steinfeld,}
\author[b]{R. Gaitskell,}
\author[a]{W. Lorenzon}

\affiliation[a]{University of Michigan, Ann Arbor, MI 48104, USA}
\affiliation[b]{Brown University, Providence, RI 02912, USA}

\emailAdd{amarascs@umich.edu}

\abstract{Measuring the scintillation and ionization yields of liquid xenon in response to ultra-low energy nuclear recoil events is necessary to increase the sensitivity of liquid xenon experiments to light dark matter.
Neutron capture on xenon can be used to produce nuclear recoil events with energies below $0.3$~keV$_\text{NR}$ via the asymmetric emission of $\gamma$ rays during nuclear de-excitation.
The feasibility of an ultra-low energy nuclear recoil measurement using neutron capture was investigated for the Michigan Xenon (MiX) detector, a small dual-phase xenon time projection chamber that is optimized for a high scintillation gain.
Simulations of the MiX detector, a partial neutron moderator, and a pulsed neutron generator indicate that a population of neutron capture events can be isolated from neutron scattering events. 
Further, the rate of neutron captures in the MiX detector was optimized by varying the thickness of the partial neutron moderator, neutron pulse width, and neutron pulse frequency.
}


\proceeding{Light Detection in Noble Elements (LIDINE)\\
  21-23 September 2022\\
  Warsaw, Poland}

\begin{document}
\maketitle
\flushbottom

\section{Introduction}
\label{sec:intro}

Lowering the nuclear recoil (NR) energy threshold of dual-phase liquid xenon (LXe) time projection chambers (TPCs) is crucial to improve the sensitivity to light Weakly Interacting Massive Particles (WIMPs) and Coherent Electron $\nu$-Nucleus Scattering (CE$\nu$NS) events.
At present the scintillation and ionization yields have been measured down to $0.45$~keV$_\text{NR}$ and $0.3$~keV$_\text{NR}$, respectively, using nuclear recoil events created by neutrons scattering off xenon \cite{lux2022improved,lenardo2019low}.

This works presents a strategy using neutron capture on xenon to identify NR events below $0.3$~keV$_\text{NR}$ that are isolated in time from NR events produced by neutron scattering.  
Simulations were run with a model of the MiX detector to quantify the rates of the neutron capture NR events that can be used to perform a measurement of the scintillation and ionization yields. 
A neutron generator capable of creating a pulsed beam of about 10~$\mu$s width, like the Adelphi Deuterium-Deuterium (DD) generator used in LZ and LUX NR calibrations \cite{lux2022improved, lz2022}, and a water tank surrounding the MiX cryostat to moderate neutrons, are crucial to produce the neutron capture NR events of interest.

\begin{figure}
  \centering
  \includegraphics[width=0.66\textwidth]{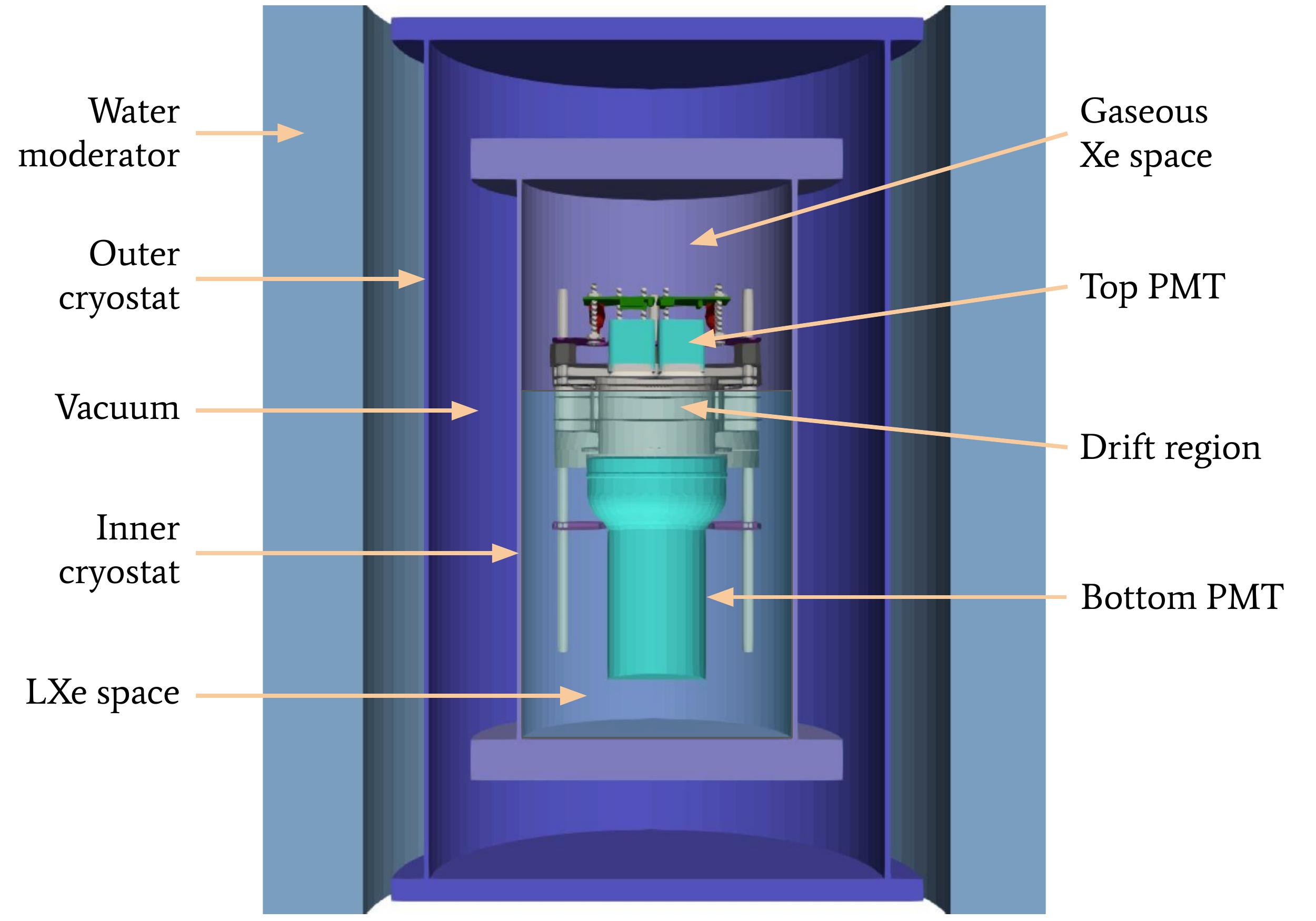}
  \caption{3D model of the MiX detector. 
  The inner cryostat encloses the LXe space that partially submerges the TPC assembly, and thus the TPC contains only a small fraction of the LXe in the system. The thickness of the water tank shown here is 5\,cm.
  }
  \label{fig:MiX}
\end{figure}

The MiX detector is a small, dual-phase, xenon time projection chamber (TPC) at the University of Michigan \cite{stephenson2015mix}.
It has a drift region with a diameter of 62.5~mm and a height of 12~mm. 
It was designed and built to have high scintillation and ionization gains, measured to be (0.239~$\pm$~0.012)~photoelectrons/photon and (16.1~$\pm$~0.6)~photoelectrons/electron, respectively. 
The MiX detector has a large volume (around 5~kg) of LXe outside the drift region, which is currently being instrumented with silicon photo-multipliers (SiPMs) to detect $\gamma$ rays from neutron capture events inside the TPC, providing a coincidence tag. 
A 3D model of the MiX detector is shown in Figure~\ref{fig:MiX}.

\section{General strategy}

Upon capturing a neutron, a xenon nucleus de-excites to the ground state within 1~ns by emitting a cascade of $\gamma$ rays, with total energy equal to the neutron separation energy $S_\text{n}$ of the newly created isotope.
The emission of the cascade imparts some recoil energy to the nucleus less than the maximum possible recoil energy
\begin{equation}
  E_\text{R, max} = \frac{S^2_\text{n}}{2M_\text{Xe}} \approx S^2_\text{n} \left (  \frac{4 \times 10^{-6}}{\text{MeV}} \right ),
\end{equation}
where $M_\text{Xe}$ is the mass of the new isotope. 
Numerical values for $E_\text{R, max}$ are given in Table. \ref{tab:nGamma} for all the naturally occurring isotopes, along with their natural abundances and thermal neutron capture cross sections. 

\begin{table}[htbp]
  \centering
  \caption{Properties of xenon nuclei that are relevant to interactions with slow neutrons: natural abundances~\cite{Gammabook}, thermal neutron capture cross sections, and the maximum recoil energy $E_\text{R, \text{max}}$ imparted to the product nuclei by the $\gamma$ cascades following capture~\cite{Gammabook}. 
  Of primary interest to the proposed measurement are $^{129}$Xe and $^{131}$Xe due to their large natural abundances, large thermal neutron capture cross sections, and the prompt $\gamma$ cascades of their capture products.
  The isotopes with missing data in the last column produce activated products upon neutron capture that do not decay promptly.}
  \smallskip
  \begin{tabular}{| c c c | c c |}
    \hline
      Target Isotope & Abundance (\%) & Cross Section (b) & Product Isotope & $E_\text{R, \text{max}}$ (keV$_\text{NR}$) \\
      \hline
      $^{124}$Xe & 0.1   & 165 $\pm$ 20    & $^{125}$Xe & 0.230 \\
      $^{126}$Xe & 0.1   & 3.8 $\pm$ 0.5   & $^{127}$Xe & - \\
      $^{128}$Xe & 1.9   & 5.2 $\pm$ 1.3   & $^{129}$Xe & 0.187 \\
      $^{129}$Xe & 26.4  & 21 $\pm$ 5      & $^{130}$Xe & 0.332 \\
      $^{130}$Xe & 4.1   & 4.8 $\pm$ 1.2   & $^{131}$Xe & 0.168 \\
      $^{131}$Xe & 21.2  & 85 $\pm$ 10     & $^{132}$Xe & 0.305 \\
      $^{132}$Xe & 26.9  & 0.42 $\pm$ 0.05 & $^{133}$Xe & - \\
      $^{134}$Xe & 10.4  & 0.27 $\pm$ 0.02 & $^{135}$Xe & - \\
      $^{136}$Xe & 8.9   & 0.26 $\pm$ 0.02 & $^{137}$Xe & 0.060 \\
      \hline
  \end{tabular}
  \label{tab:nGamma}
\end{table} 

\begin{figure}
  \centering
  \includegraphics[width=0.55\textwidth]{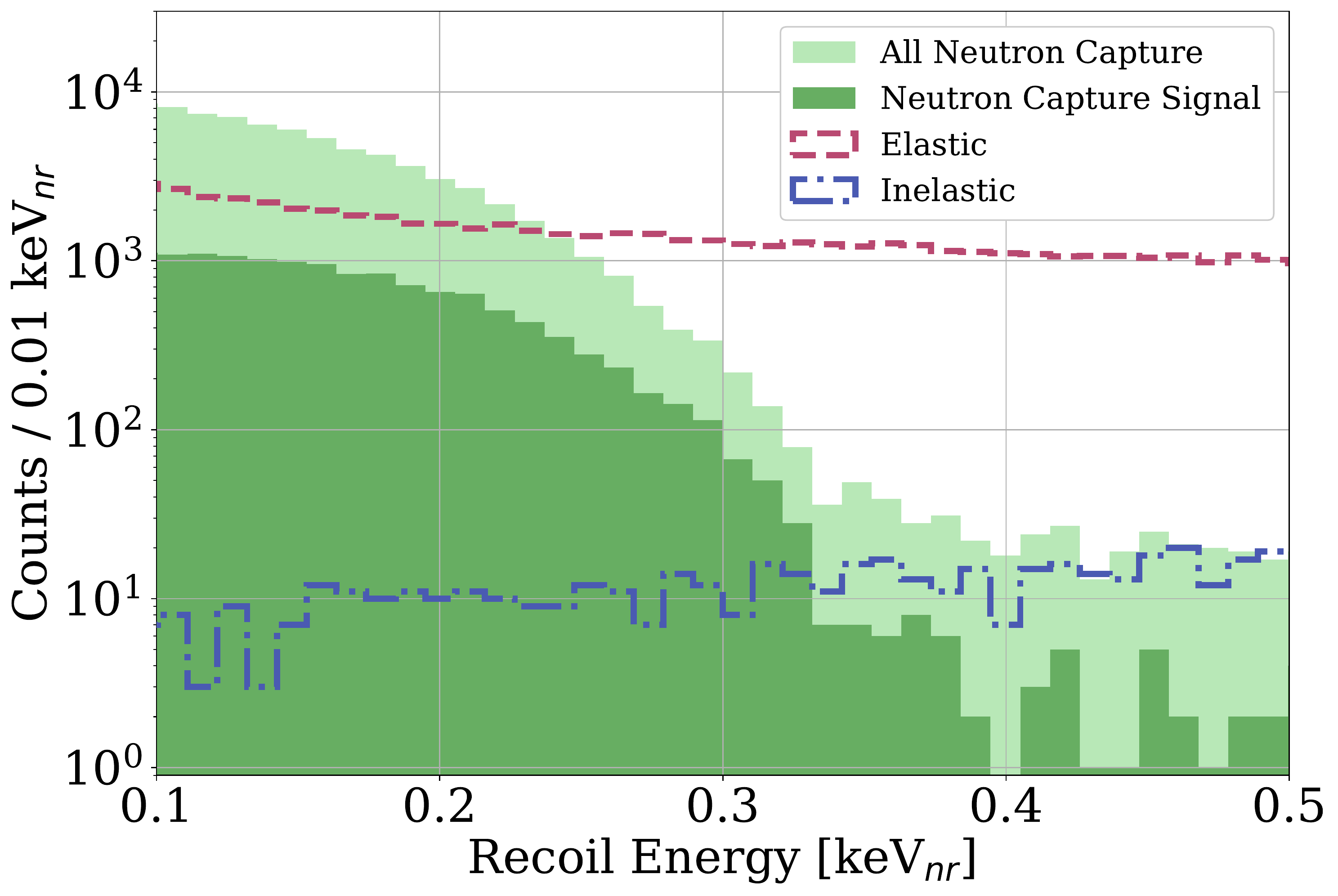}
  \caption{Nuclear recoil spectrum due to neutron interactions simulated in the MiX detector. 
  The shaded light green histogram (140,000 counts) shows all recoil events due to neutron captures, while the shaded dark green portion (20,000 counts) only retains those where all of the $\gamma$-rays from the nuclear de-excitation process escape the active volume.
  Also shown are the recoil events due to elastic (dashed magenta) and inelastic (dashed-dot blue) neutron scatters. 
  All inelastic scatters are shown, regardless of whether their de-excitation $\gamma$-products escape the TPC.}
  \label{fig:n capture}
\end{figure}

GEANT4 simulations were performed using the QGSP\_BIC\_HP physics list for neutron transport, and the photon evaporation model was used to model the nuclear de-excitation process.
Of all neutron capture events in the MiX TPC, only events where the entire $\gamma$ cascade escapes the TPC, called \emph{signal events}, can be used for the measurement.
Figure.~\ref{fig:n capture} shows a simulated NR spectrum and the subset of signal events in the MiX TPC when a point source of 2.45~MeV neutrons (as if emitted from the DD generator) 1~m away from the TPC is exposed to the detector surrounded by a 5~cm thick water tank. 
According to the simulation, 15\% of neutron captures in the TPC emit $\gamma$ cascades that escape the TPC.

This measurement relies on the GEANT4 model of energy deposition due to neutron capture.
The distribution of simulated NR energies will be used to calculate the sizes of the scintillation and ionization signals according to parameterized estimates of the yields below $0.3$ keV$_\text{NR}$.
These parameters can be adjusted to fit the calculated signals to the observed data, as performed in Ref.~\cite{dqthesis}.

\subsection{Time structure of neutron interactions}

\begin{figure}
  \centering
  \includegraphics[width=0.55\textwidth]{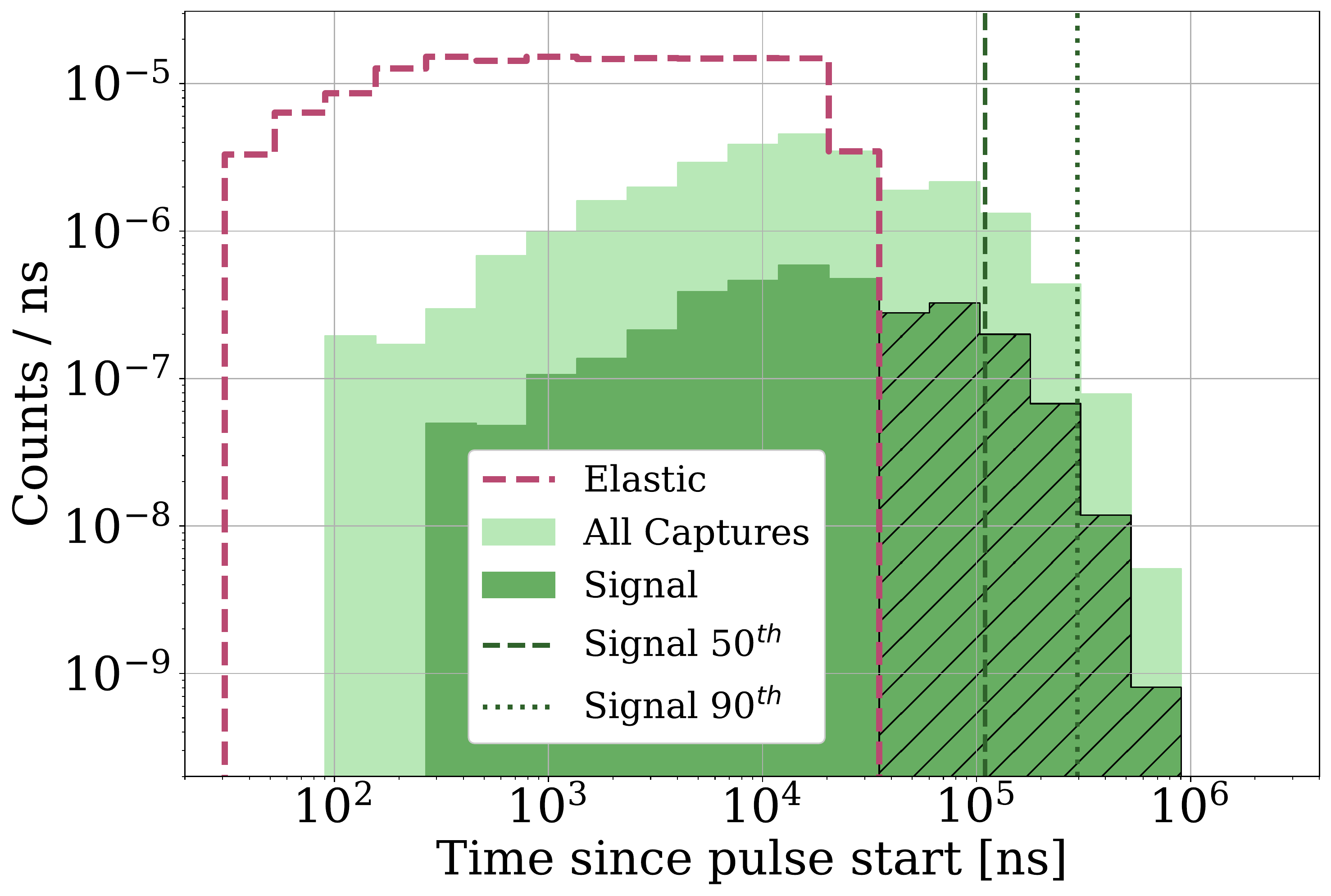}
  \caption{Time distribution of the neutrons that interact with the active LXe volume in the TPC, from a simulation done for a 5\,cm water tank and $w_n = 30\,\mu$s.
  The total counts due to neutron capture (light green) and elastic scattering (dashed magenta) are normalized to unity.
  Inelastic scattering events are omitted from this plot for clarity as their rate is a hundred-fold less than the elastic rate.
  All events shown here deposit less than $1$ keV$_\text{NR}$.
  The dark green histogram shows all signal events, and the hatched portion shows the signal events that occur after the last scattering time, i.e., in the signal window.
  Visual checkpoints for when $50\%$ and $90\%$ of all signal events occur are shown with the vertical dashed and dotted lines, respectively.
  }
  \label{fig:timeStructure}
\end{figure}

When a pulsed beam of neutrons is shot at the detector, a set of neutron capture events suitable for measuring yields is produced in each pulse. 
Since the neutron capture cross section is roughly proportional to the inverse speed of the incident neutron (except at resonances), neutron capture events are produced by slow neutrons in the detector. 
The role of the moderator is to slow down the monoenergetic neutrons from the source, while discouraging capture in the moderator material itself, as the resulting $\gamma$ rays create pile-up.
Accordingly, the simulation shows that partial neutron moderation is ideal. 
The moderator allows fast neutrons into the detector first, which are likely to scatter, followed by slower neutrons that are captured. 
In this arrangement, NR events due to neutron capture can be isolated from scattering events with an appropriate time cut. 
The width of the neutron pulse $w_n$ and the thickness of the water tank can be chosen to optimize the number of signal events surviving the time cut.
Figure~\ref{fig:timeStructure} shows the time distributions of neutron interactions in the TPC for $w_n = 30$~$\mu$s and a water tank thickness of 5~cm.
The period of time after the elastic scattering rate has diminished, and before 99\% of neutron capture events occur, is called the \emph{signal window}.

Signal events can be positively identified if their $\gamma$ cascades are detected outside the TPC.
In the MiX detector, a natural location to detect the interactions of these $\gamma$ rays is the large volume of LXe immediately outside the TPC.
Simulations show that 70\% of signal events can be tagged using an instrumented skin with a 100\,keV$_\text{ee}$ energy threshold. 
In other words, less than 30\% of signal events emit $\gamma$ cascades that escape not only the TPC, but the surrounding skin region as well.
No significant bias on the NR energy spectrum is observed when taggable signals are selected. 
Tagging offers a major reduction in the single-electron background commonly observed in LXe TPCs that may otherwise dominate the number of events from neutron capture that also produce single electrons~\cite{akerib2020investigation}.

\subsection{Neutron capture model uncertainty}

\begin{figure}
  \centering
    \includegraphics[width=0.58\textwidth]{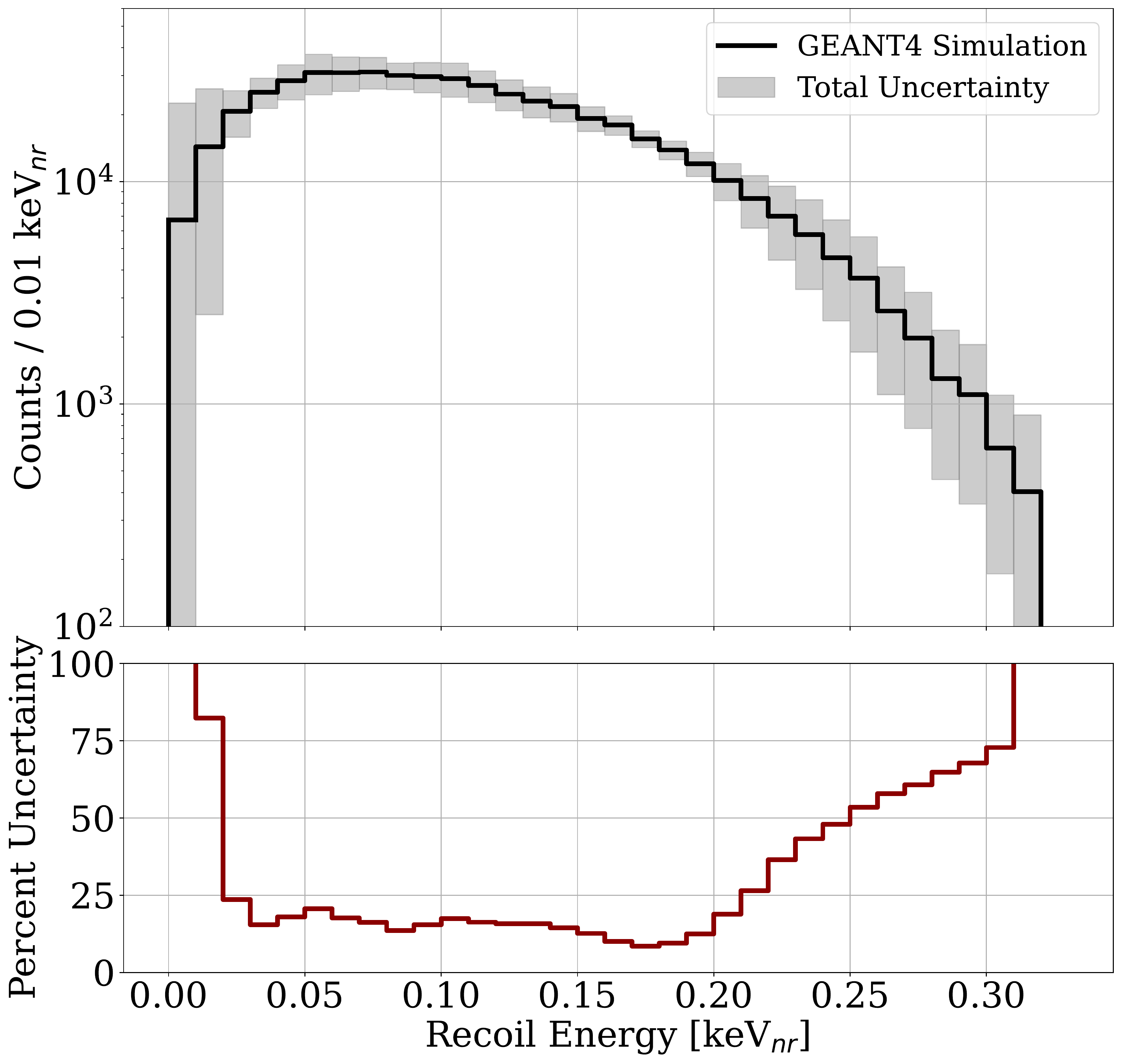}
    \caption{\emph{Top:} NR spectrum due to thermal neutron capture in LXe simulated using GEANT4. The gray uncertainty band represents the total uncertainty, which incorporates discrepancies in the $\gamma$ spectra between the ENSDF and the EGAF files.
    \emph{Bottom:} The error band in the top panel is presented as percent uncertainty for clarity.
    }
    \label{fig:nCapError}
\end{figure}

Since the experimental concept relies on a comparison with simulations, a custom algorithm was implemented to generate nuclear recoils from neutron capture and used to calculate the uncertainty of the NR energy spectrum.
This uncertainty, {shown in Fig.~\ref{fig:nCapError}, incorporates discrepancies in the $\gamma$ spectra between ENSDF and the evaluated gamma ray activation file (EGAF)~\cite{EGAFdata}}, in addition to a conservative range of multiplicity distributions.
Details of the uncertainty calculation are given in the appendix of Ref.~\cite{amarasinghe2022feasibility}.

It is apparent that comprehensive nuclear data are required to constrain the uncertainty in the NR spectrum.
A further improvement to the energy deposition model can be made by incorporating angular correlations of $\gamma$ rays in the cascade.
Measurements of the angular correlations have been made for prominent transitions in the $^{130}$Xe and $^{132}$Xe cascades \cite{hamada1988gamma}, but are not yet taken into account in our simulations.  

\section{Expected backgrounds}
\label{sec:backgrounds}

There are three types of non-NR events that could reduce the rate of signal events:
i) the low energy electronic recoil (ER) background from the $\gamma$ cascades of activated and capture products, and from radiation in the environment,
ii) the single electron (SE) background, and
iii) the high energy ER events in the TPC that could temporarily blind the detector. 
The simulation indicates that the rate of low energy ER events produced by $\gamma$ cascades from neutron captures outside the TPC is at least an order of magnitude lower than the rate of signal events, so they are not considered further.

\subsection{Single electron background}
Single electron backgrounds are one of the biggest obstacles to the low energy sensitivity of LXe TPCs, but cannot be reliably simulated as they depend on the various runtime parameters, e.g. LXe purity, drift field, and extraction field.
High rates of single electrons have been observed after large ionization signals by the LUX collaboration, and they decay over timescales of about 10~ms \cite{akerib2020investigation}.
The capture events following a neutron pulse produce a $\gamma$-rich environment in which many large ionization signals are expected. 
This information was used along with the measured decay time to set a conservatively low neutron pulsing frequency of 60~Hz.
This value was obtained by trading off the large number of signal events provided by a high frequency and the correspondingly high rate of single electron background.
A further optimization can be done following a characterization of the single electron background in the MiX detector.
A significantly larger pulsing frequency is allowed with the tagging capability provided by the instrumented volume of LXe outside the TPC, and shall be quantified following the measurement of the single electron background.

\subsection{Pileup}
\begin{figure}
  \centering
  \includegraphics[width=0.66\textwidth]{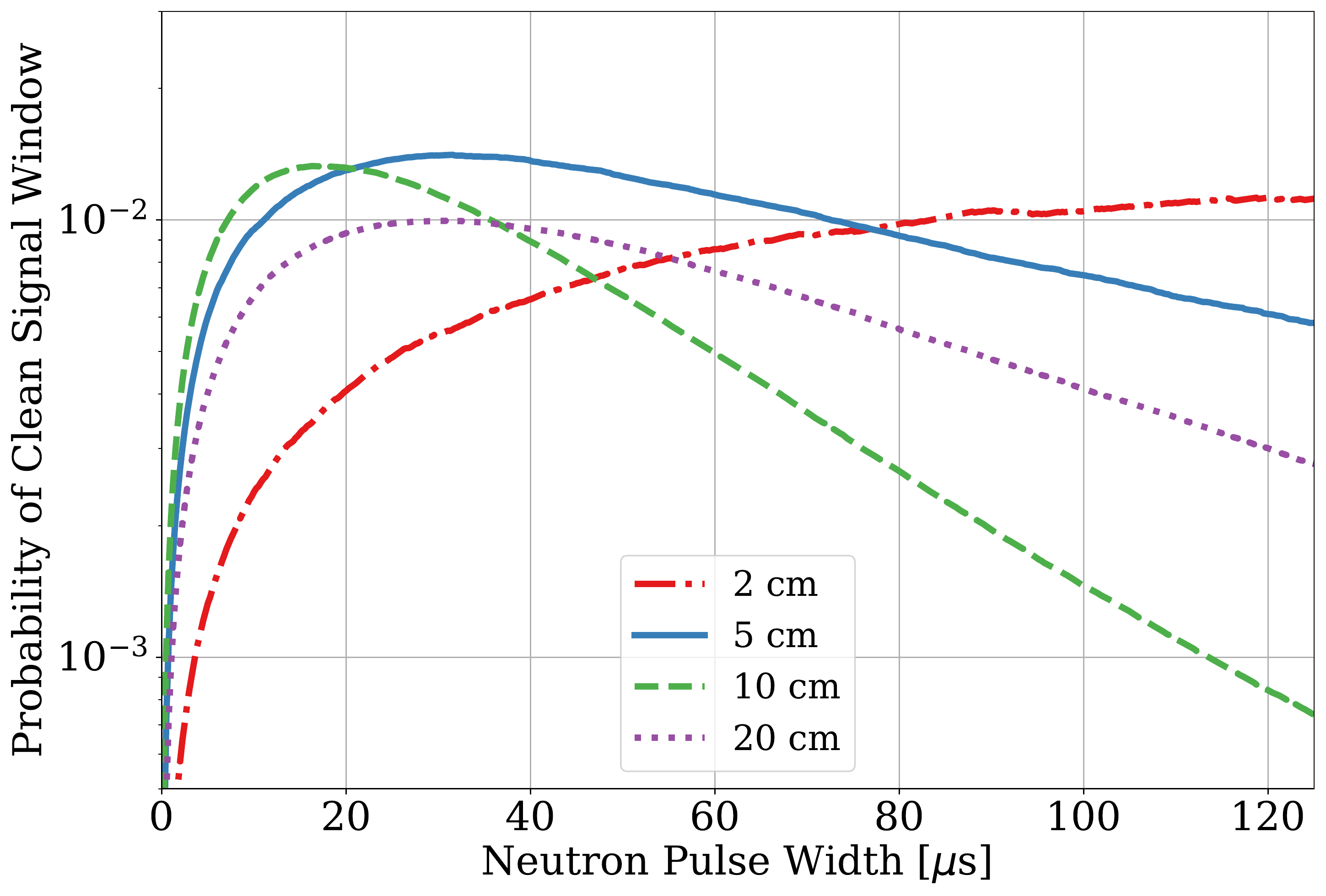}
  \caption{Probability $P$ of obtaining a clean signal window where no signal event is accompanied by an ER deposit, as a function of neutron pulse width $w_n$. 
  Curves are shown for a representative set of water tank thicknesses.
  }
  \label{fig:cleanSignal}
\end{figure}

High energy ER events produced by the $\gamma$-rich environment can coincide with the signal events and contribute to pileup, decreasing the number of clean acquisition windows that contain only the scintillation and ionization pulses of the neutron capture event.
This pileup provides the strongest constraints to the water tank thickness and neutron pulse width. 
The effect of pileup is quantified by the probability $P$ that a given signal window has no large ER deposit, while containing a signal event, such that
\begin{equation}
  P = \text{Pois} (0, \text{ER}_\text{ext}) \times \text{Pois} (1, \text{NR}_\text{signal}),
\end{equation}
where ER$_\text{ext}$ is the average number of ER events in a signal window produced by the $\gamma$ cascades of neutron captures outside the TPC, and NR$_\text{signal}$ is the average number of signal events in a signal window.  
Figure~\ref{fig:cleanSignal} shows the probability $P$ as a function of $w_n$ for a set of water tank thicknesses. 
A 5~cm thickness and a pulse width of 30~$\mu$s is identified as optimal.

\section{Summary}

The focus of this study is an experimental concept to measure the NR scintillation and ionization yields of LXe below $0.3$~keV$_\text{NR}$. 
GEANT4 simulations of the MiX detector, a pulsed neutron source, and a thin neutron moderator were done to establish the feasibility to use neutron capture events as a source for this measurement.
Pileup events and single electron backgrounds were taken into account to maximize the rate of neutron capture signal events. 
The best configuration is a water tank thickness of 5~cm, neutron pulse width of 30~$\mu$s, and a neutron pulsing frequency of 60~Hz, corresponding to a signal rate of 0.2 events per second.
Assuming the extrapolation of the yields below 0.3~keV$_\text{NR}$ by NEST v2.0.1 is true, a two-month run with the MiX detector will provide enough data to measure the yields down to 0.13~keV$_\text{NR}$. 

\acknowledgments

We thank M. Solmaz, M. Szydagis, P. Sorensen, R. Linehan, J. Liao, and X. Xiang for valuable conversations.
This work was supported by the U.S. Department of Energy under Grant SC0019193, and with resources provided by the Open Science Grid \cite{pordes2007open,sfiligoi2009pilot}, which is supported by the National Science Foundation award \#2030508. 

\bibliographystyle{JHEP}

\providecommand{\href}[2]{#2}\begingroup\raggedright\endgroup


\end{document}